\documentclass[twocolumn]{article}

\usepackage[colorlinks]{hyperref} 

\usepackage{color,amssymb,mathptm,graphicx}

\newcommand{\mybox}[3]
{ 
  \put(#1,#2){\framebox(1,1){#3}}
}

\include{Qcircuit}

\title{Projecting onto Qubit Irreps of Young Diagrams}

\author{Stephen S. Bullock\footnote{IDA Center for Computing Sciences,
17100 Science Drive, Bowie, MD 20715-4300 USA, ssbullo@super.org}}

\date{June 20$^{\mbox{\footnotesize th}}$, 2006}

\begin{document}

\maketitle

\begin{abstract}
Let $K$ be the diagonal subgroup of $U(2)^{\otimes n}$.
For the one-qubit state space $\mathcal{H}_1=\mathbb{C}\{\ket{0}\}
\oplus \mathbb{C} \{ \ket{1}\}$, we may view $\mathcal{H}_1$
as a standard representation of $U(2)$ and 
the $n$-qubit state space $\mathcal{H}_n=(\mathcal{H}_1)^{\otimes n}$
as the $n$-fold tensor product of standard representations.  Representation
theory then decomposes $\mathcal{H}_n$ into irreducible
subrepresentations of $K$
parametrized by combinatorial objects known as \emph{Young diagrams}.
We argue that $n-1$ classically controlled measurement circuits,
each a Fredkin interferometer, may be used to form a projection operator
onto a random Young diagram irrep within $\mathcal{H}_n$.  For
$\mathcal{H}_2$, the two irreps happen to be orthogonal and correspond
to the symmetric and wedge product.  The latter
is spanned by $\ket{\Psi^-}$, and the standard two-qubit
swap interferometer requiring a single Fredkin gate suffices in this case.
In the $n$-qubit case, it is possible to extract many copies of
$\ket{\Psi^-}$.  Thus applying this process using nondestructive
Fredkin interferometers allows for the creation of entangled bits
(e-bits) using fully mixed states and von Neumann measurements.
\end{abstract}

\section{Introduction}

Theoretical quantum computing considers data to be stored within 
idealized quantum particles and then makes inferences on how such 
data may be manipulated in terms of the axioms of quantum mechanics.
Since quantum measurement changes the state of the underlying particles,
it is not merely a matter of input/output but rather a computational act.
Consider for example how crucial the appropriate measurements are for
quantum teleportation.  More generally, a thread of recent research
demonstrates that any quantum circuit may be emulated by a chain
of carefully orchestrated measurements on a large highly entangled
quantum register (e.g. \cite{Jozsa}.)

This work is much more modest, in that it presents a sequence of
classically controlled quantum circuits realizing a projector
which seems to have been overlooked.  The observation is related to
but simpler than work of Bacon, Chuang, and Harrow on
Clebsch-Gordon transforms \cite{Harrow1,Harrow2}.  Given
$\mathcal{H}_1=\mathbb{C} \{\ket{0}\} \oplus \mathbb{C}\{ \ket{1}\}$
and $\mathcal{H}_n=\mathcal{H}_1^{\otimes n}$ the $n$-qubit state
space, both the Clebsch-Gordon transformation and this work make essential
use of the (classical) representation theory of $\mathcal{H}_n$.
For we may view $\mathcal{H}_1$ as the standard representation of
$U(2)$, so that $\mathcal{H}_n$ is the $n$-fold tensor product of
standard representations equipped with the left-multiplication 
by elements of $K$, the diagonal subgroup of $U(2)$:
\begin{equation}
K \ {\buildrel \mbox{\tiny def } \over =} \ 
\{ \; V^{\otimes n} \; ; \; V \in U(2) \; \}
\end{equation}
Also, $\mathcal{H}_n$ inherits a standard representation
by $S_n$, the group of permutations of the elements of the
set $\{1,2,\ldots,n\}$.  Namely, if $\sigma \in S_n$, then
we act by the permutation unitary $U_\sigma$ which satisfies
$U_\sigma \ket{b_1 b_2 \ldots b_n}=\ket{b_{\sigma(1)} \ldots
b_{\sigma(n)}}$ on all computational basis kets.  Then
$k U_\sigma \ket{\psi} = U_\sigma k \ket{\psi}$ for all
$\sigma \in S_n$, $k \in K$.  A \emph{subrepresentation} w.r.t. $K$
is a linear subspace $\mathcal{L} \subseteq \mathcal{H}_n$
preserved by all $k \in K$.  The definition extends to considering
subrepresentations of subrepresentations, and a subrepresentation
$\mathcal{L}$ is \emph{irreducible} if the only subrepresentations
of $\mathcal{L}$ are $\{ 0 \}$ and $\mathcal{L}$. 
Now suppose any decomposition of $\mathcal{H}_n$ into
irreducible subrepresentations (henceforth irreps) of $K$ above.
Each element $\sigma \in S_n$
will permute the factors, since $\sigma(\mathcal{L})$ is another
irrep isomorphic to the original under $\sigma^{-1}$
and hence irreducible.

\noindent
{\bf Example 1:}  $\mathcal{L}=\mathbb{C}\{ \ket{\psi_1}= 
\ket{0011}-\ket{1001}-\ket{0110}+\ket{1100} \}$
is a subrepresentation of $\mathcal{H}_4$, since the ket
lies within a decoherence-free-subspace (DFS) 
\cite{dfsi,dfsii,dfsiii}
on which each $k \in K$
satisfies $k \ket{\psi_1}= \mbox{det}(k)^2\ket{\psi_1}$.  Also,
any one-dimensional subrepresentation is irreducible.  If $(23)$
denotes the flip permutation exchanging $2$ and $3$, then
$U_{(23)} \mathcal{L}= \mathbb{C}
\{\ket{0101}-\ket{1001}-\ket{0110}+\ket{1010} \}$ 
spans a second dimension of the DFS.

Earlier works on quantum circuits for $K$-irreps \cite{Harrow1,Harrow2}
describe input-output register indexing schemes for the Clebsch-Gordon
transform, which in particular carries each computational basis state
into an irrep of $\mathcal{H}_n$ indexed by a
combinatorial object known as a Young diagram.  The second work
applies equally well to qubits and qudits.  In contrast, this work
attempts only gadgetry: a scheme exploiting classical control is outlined
for projecting onto Young diagram irreps.  Before outlining the
general case in the next section, we argue that a Fredkin interferometer
suffices in two qubits.  The decomposition into irreps is as follows,
where we have placed the appropriate Young diagram below each irrep.
\begin{equation}
\begin{array}{lcccc}
\mathcal{H}_2 & = & \mbox{Sym}^2(\mathcal{H}_1) & \oplus 
& \wedge^2 (\mathcal{H}_1)
\\
& = & 
\begin{picture}(2,1)
\mybox{0}{0}{1}
\mybox{1}{0}{2}
\end{picture}
& {\buildrel \bigoplus \over \ } &
\begin{picture}(1,2)
\mybox{0}{1}{1}
\mybox{0}{0}{2}
\end{picture} \\
\end{array}
\end{equation}
Here, $\mbox{Sym}^2(\mathcal{H}_1)=\mbox{span}_{\mathbb{C}}
\{ \ket{00},\ket{01}+\ket{10},\ket{11}\}=
\mbox{span}_{\mathbb{C}} \{ \ket{\Phi^\pm},\ket{\Psi^+}\}$ is the 
symmetric irrep of $K$ consisting of states invariant under
SWAP.  The other irrep is anti-invariant under SWAP.
Setting notation, let $\ket{\Psi^-}=2^{-1/2}(\ket{01}-\ket{10})$,
so that $\wedge^2(\mathcal{H}_1) = \mathbb{C} \{ \ket{\Psi^-}\}$.  
For $(12)$ the nontrivial element of $S_2$, we have $U_{(12)}=\chi$
(the SWAP operator) and $\chi \ket{\Psi^-}=-\ket{\Psi^-}$.
Finally, consider the following quantum logic circuit, in this work
described as a Fredkin gate interferometer.
\begin{center}
\[
\Qcircuit @R=2.0em @C=.55em  {
& \lstick{\ket{0}} & \qw & \gate{H} & \ctrl{1} 
& \gate{H} & \qw & \meter & \cw \\
&  & \qw & \qw & \qswap \qwx & \qw & \qw & \qw & \qw \\
&  & \qw & \qw & \qswap \qwx & \qw & \qw & \qw & \qw }
\]
\end{center}
Here, $H=2^{-1/2} \sum_{b_1,b_2=0}^1 (-1)^{b_1 b_2} \ket{b_2} \bra{b_1}$
is the one-qubit Hadamard unitary.
By a direct computation or due to the interferometer-circuit
literature \cite{LLW,Brun,Ekert}, when $\ket{\psi}$ is placed
on the input lines the measurement satisfies
\begin{equation}
\mbox{Prob}(\mbox{ancilla}=0)=\frac{1}{2}+{\frac{1}{2}}
\mbox{Trace}(\chi \ket{\psi}\bra{\psi})
\end{equation}
Hence the two outputs of the Fredkin interferometer perform
projections onto $\mbox{Sym}^2(\mathcal{H}_1)$ 
and $\wedge^2 (\mathcal{H}_1)$.  This work extends
the above Young diagram projector into $n$ qubits.

An application of a sort results.  Recall that entangled-bits
(or e-bits) are fully-entangled, pure two-qubit states.
All such states are equivalent under the action of some
$U_1 \otimes U_2$ for $U_1$, $U_2$ being two-by-two unitary
matrices.  For example, if the usual e-bit is
$\ket{\Phi^+}=2^{-1/2}(\ket{00}+\ket{11})$
and $\ket{\Psi^-}=2^{-1/2}(\ket{01}-\ket{10})$,
then $\sigma_x \sigma_z \otimes I_2 \ket{\Psi^-}=\ket{\Phi^+}$.
Such e-bits are required for quantum teleportation and are
fundamental resources in theoretical works generalizing
Shannon's theory of classical channel capacities to
quantum channels \cite{q_Shannon}.

As is well-known, an e-bit may be produced by a unitary process.
For $\ket{\Phi^+}$ results from the following diagram.
\begin{center}
\[
\Qcircuit @R=2.0em @C=.55em  {
& \lstick{\ket{0}} & \qw & \gate{H} & \ctrl{1} & \qw \\
& \lstick{\ket{0}} & \qw & \qw & \targ & \qw
}
\]
\end{center}
Also, e-bits may be produced by cooling,
since the Hamiltonian $H_0=-\sigma_z \otimes \sigma_z$ holds
$\ket{\Phi^+}$ within its groundstate and any
small perturbation $H_\epsilon=-\sigma_z \otimes \sigma_z - \epsilon
\sigma_x \otimes \sigma_x$ will split the degeneracy
so that $\ket{\Phi^+}$ is the unique groundstate.  A third option
is to prepare an e-bit using a von Neumann measurement within the
Fredkin gate interferometer, as above.  The yield would seem
to be low.  For say the input is a fully mixed state
$I_4/4$.  Then the probability of creating
$\ket{\Psi^-}$ is only $1/4$, since $\mbox{dim } \wedge^2(\mathcal{H}_1)=1$
and $\mbox{dim } \mbox{Sym}^2(\mathcal{H}_1)=3$.  Our generalization
to $n$-qubit Young diagrams shows that much more efficient initialization
of e-bits using von Neumann measurements is possible for $n>2$.

This manuscript is organized as follows.
In \S \ref{sec:young}, we review the irreps of the two-qubit state
space and their Young diagrams.
In \S \ref{sec:projectors}, we present an algorithm for choosing a
sequence of (classically-controlled) von Neumann measurements.
In \S \ref{sec:ebits}, we present an application, namely
harvesting copies of $\wedge^2(\mathcal{H}_1)$ as e-bits.

\section{Young diagram irreps}

\label{sec:young}

\setlength{\unitlength}{0.25in}

The representation of theory of $\mathcal{H}_n$ under either $K$
or the symmetric group on $\{1,2,\ldots,n\}$ is both classical
and well-known \cite{Goodman}.
To begin, it suffices to produce a highest-weight $\ket{\psi}$ within 
each irrep.  For given such a $\ket{\psi}$, it happens that
$\tilde{X}^j \ket{\psi}$ span the irrep, where
$\tilde{X}=\sum_{k=1}^n X_k \in i \mathfrak{k}$ is a local Hamiltonian.
Each (standard) Young diagram specifies such a highest-weight
vector.  For a two-level
system, a diagram consists of two rows of $n$ boxes total.  
Each box holds an integer within $\{ 1,2,\ldots,n\}$,
no integers are repeated, and the integers increase moving down each
column and across both rows.  We use the following notation, where
$p+q=n$.
\begin{equation}
\label{eq:tableaux}
\begin{picture}(8,2)
\mybox{0}{0}{$k_1$}
\mybox{1}{0}{$k_2$}
\mybox{2}{0}{$\cdots$}
\mybox{3}{0}{$k_q$}
\mybox{0}{1}{$j_1$}
\mybox{1}{1}{$j_2$}
\mybox{2}{1}{$\cdots$}
\mybox{3}{1}{$j_q$}
\mybox{4}{1}{$j_{q+1}$}
\mybox{5}{1}{$\cdots$}
\mybox{6}{1}{$j_{p-1}$}
\mybox{7}{1}{$j_p$}
\end{picture}
\end{equation}
For any Young diagram $p \geq q$, $j_1=1$, and $k_q=n$.  Let $G$
denote the group of all permutations of $\{1,2,\ldots,n\}$ which
preserve the subsets corresponding to the columns of the
Young diagram.  Let $b(\ell)$ denote the bit which is $0$ if 
$\ell \in \{1,2,\ldots,n\}$
occurs in the top row and $1$ if $\ell$ occurs in the bottom row.
Then the highest-weight vector corresponding to the diagram $\mathcal{T}$ is
\begin{equation}
\ket{\psi_{\mathcal{T}}} \ = \ 
(\# G)^{-1/2} \sum_{\sigma \in G} \mbox{sign}(\sigma)
U_\sigma \ket{b(1) b(2) \ldots b(n)}.
\end{equation}
If $\mathcal{L}_{\mathcal{T}}$ is the corresponding irrep, 
then the basic result of classical representation
theory states (\cite{Goodman,Harrow2})
$\mathcal{H}_n=\oplus \mathcal{L}_{\mathcal{T}}$.  Here, the
\emph{nonorthogonal} (vector-space) direct sum is taken over all
Young diagrams.

\setlength{\unitlength}{0.125in}

\noindent
{\bf Example 1 (Cont.):}  Consider the following four-qubit Young diagram.
\begin{center}
\begin{picture}(2,2)
\mybox{0}{0}{2}
\mybox{1}{0}{4}
\mybox{0}{1}{1}
\mybox{1}{1}{3}
\end{picture}
\end{center}
Then $G=\{(),(12),(34),(12)(34)\}$ so that
\begin{equation}
\begin{array}{lcl}
\ket{\psi_{\mathcal{T}}}
& =  &  (1/2)(I_4-U_{(12)}-U_{(34)}+U_{(12)(34)}) \ket{0101} \\
&  = & (1/2)(\ket{0101}-\ket{1001}-\ket{0110}+\ket{1010} ) \\
\end{array}
\end{equation}
Thus the trivial irrep associated to the four-qubit DFS may be
derived from the Young diagram above.  Applying $U_{(23)}$ to the
above ket produces a ket corresponding to
\begin{center}
\begin{picture}(2,2)
\mybox{0}{0}{3}
\mybox{1}{0}{4}
\mybox{0}{1}{1}
\mybox{1}{1}{2}
\end{picture}
\end{center}
Thus diagram irreps need not be orthogonal, given the earlier discussion.
The $\Gamma$-shaped diagram in three-qubits provide another example.

\noindent
{\bf Example 2:}  Consider the horizontal Young diagram
for which $q=0$.  Then we iteratively apply $\tilde{X}$ to
$\ket{00\ldots0}$.  Now $\tilde{X} \ket{00\ldots 0}$ is a singlet,
$\tilde{X}^2\ket{00\ldots0}$ is a doublet, etc.  The resulting irrep
is $\mbox{Sym}^n(\mathcal{H}_1)$, i.e. the $K$-irrep consisting
of those kets invariant under all $U_\sigma$ for $\sigma \in S_n$.

\section{Irrep projectors}

\label{sec:projectors}

\setlength{\unitlength}{0.25in}

When taking tensor products of irrep subspaces as specified by
Young diagrams, one forms a direct sum of all possible concatenations
of the diagrams which are themselves (standard) Young diagrams.  The
resulting direct summands are often orthogonal, since nonorthogonal
Young diagram irreps must have the same shape.  In symbols, suppose
diagrams $\mathcal{T}_1$ and $\mathcal{T}_2$ with corresponding
$p_1,q_1$ and $p_2,q_2$ respectively per Equation \ref{eq:tableaux}.
Then Schur orthogonality demands
$\mathcal{L}_{\mathcal{T}_1} \perp \mathcal{L}_{\mathcal{T}_2}$
whenever $p_1 \neq p_2$ (equivalently $q_1 \neq q_2$.)
In this way tensor products often produce orthogonal direct sums.

The simplest example of a tensor product involves rectangular diagrams.  
Suppose a diagram according to Equation \ref{eq:tableaux}, except that
for purposes of induction say $p+q=n-1$ rather than $n$, and
let $K_{n-1}$ be the diagonal subgroup of $U(2)^{\otimes (n-1)}$ with
$K_n$ similar.  We tensor
the Young diagram irrep with $\mathcal{H}_1$ on qubit $n$, whose
corresponding Young diagram is a single box containing $n$.
Suppose further that the original diagram is rectangular,
i.e. $p=q$.  Then the only valid concatenation is that diagram
for which $j_{p+1}=n$, and the tensor product is itself this
diagram irrep on $K_n$.

Consider next a qubit irrep of a Young diagram which is not
rectangular.  Arguing as above with a one-row diagram replacing
$n$, we see that the diagram of Equation \ref{eq:tableaux} is also
the following tensor product:
\begin{equation}
\label{eq:rec_tens_line}
\begin{picture}(9,2)
\mybox{0}{0}{$k_1$}
\mybox{1}{0}{$k_2$}
\mybox{2}{0}{$\cdots$}
\mybox{3}{0}{$k_q$}
\mybox{0}{1}{$j_1$}
\mybox{1}{1}{$j_2$}
\mybox{2}{1}{$\cdots$}
\mybox{3}{1}{$j_q$}
\put(4.25,0.75){$\otimes$}
\mybox{5}{1}{$j_{q+1}$}
\mybox{6}{1}{$\cdots$}
\mybox{7}{1}{$j_{p-1}$}
\mybox{8}{1}{$j_p$}
\end{picture}
\end{equation}
The latter factor is a copy of $\mbox{Sym}^{p-q}(\mathcal{H}_1)$,
mapped into the $p-q$ qubits labelled by contents of the diagram
boxes.  The rectangular factor is a one-dimensional representation.
To see this, consider that the rectangular diagram is the only possible
concatenation of the tensor product:
\begin{equation}
\label{eq:lots_of_singlets}
\begin{picture}(12,2)
\mybox{0}{0}{$k_1$}
\mybox{1}{0}{$k_2$}
\mybox{2}{0}{$\cdots$}
\mybox{3}{0}{$k_q$}
\mybox{0}{1}{$j_1$}
\mybox{1}{1}{$j_2$}
\mybox{2}{1}{$\cdots$}
\mybox{3}{1}{$j_q$}
\put(4.25,0.75){=}
\mybox{5}{0}{$k_1$}
\mybox{5}{1}{$j_1$}
\put(6.25,0.75){$\otimes$}
\mybox{7}{0}{$k_2$}
\mybox{7}{1}{$j_2$}
\put(8.25,0.75){$\otimes$}
\put(9.25,0.75){$\cdots$}
\put(10.25,0.75){$\otimes$}
\mybox{11}{0}{$k_q$}
\mybox{11}{1}{$j_q$}
\end{picture}
\end{equation}
Each column describes a copy of $\ket{\Psi^-}$ on the appropriate bits.  
Thus, $k\in K$ acts as multiplication by $\mbox{det}(k)^q$
on the subspace spanned by qubits whose labels appear in the rectangular
subrepresentation.   We refer to the 
box containing $j_{q+1}$ of Equation \ref{eq:tableaux} as the hook-box
of a diagram which is not rectangular.

What happens when we tensor the irrep of Young diagram which
is not rectangular with $\mathcal{H}_1$?
There are only two possible concatenations appear below, and they
result by placing the box with contents $n$ below or
far to the right of the hook-box.
\setlength{\unitlength}{0.25in}
\begin{equation}
\label{eq:decomposition}
\begin{picture}(9,5)
\mybox{0}{0}{$k_1$}
\mybox{1}{0}{$k_2$}
\mybox{2}{0}{$\cdots$}
\mybox{3}{0}{$k_q$}
\mybox{4}{0}{$n$}
\mybox{0}{1}{$j_1$}
\mybox{1}{1}{$j_2$}
\mybox{2}{1}{$\cdots$}
\mybox{3}{1}{$j_q$}
\mybox{4}{1}{$j_{q+1}$}
\mybox{5}{1}{$\cdots$}
\mybox{6}{1}{$j_{p-1}$}
\mybox{7}{1}{$j_p$}
\put(4.25,2.25){$\oplus$}
\mybox{0}{3}{$k_1$}
\mybox{1}{3}{$k_2$}
\mybox{2}{3}{$\cdots$}
\mybox{3}{3}{$k_q$}
\mybox{0}{4}{$j_1$}
\mybox{1}{4}{$j_2$}
\mybox{2}{4}{$\cdots$}
\mybox{3}{4}{$j_q$}
\mybox{4}{4}{$j_{q+1}$}
\mybox{5}{4}{$\cdots$}
\mybox{6}{4}{$j_{p-1}$}
\mybox{7}{4}{$j_p$}
\mybox{8}{4}{$n$}
\end{picture}
\end{equation}
For simplicity, relabel $j_{q+1}=j$.  Due to comments of the last
paragraph, any $\ket{\psi}$ within the irrep of the top direct
summand will satisfy $U_{(jn)} \ket{\psi}=\ket{\psi}$.
On the other hand, $U_{(jn)} \tilde{X}=\tilde{X} U_{(jn)}$,
from which we may infer that any $\ket{\psi}$ within the
irrep of the second diagram will satisfy
$U_{(jn)} \ket{\psi}=-\ket{\psi}$.  For the same was true
by construction of $\ket{\psi_{\mathcal{T}_{\mbox{\tiny bottom}}}}$, the
highest weight ket of the irrep of the lower Young diagram.
Thus, the direct sum is as promised orthogonal, and each irrep
lies in the $+1$ and $-1$ eigenspace of
$U_{(jn)}$ respectively.

\begin{table}
\begin{center}
\begin{tabular}{||r|r|r||}
\hline
qubits & $\mu$ & $\sigma$ \\
\hline
\hline
2 & 1.500 & 0.866 \\ 
\hline
3 & 2.000 & 1.000 \\ 
\hline
5 & 2.750 & 1.392 \\
\hline 
10 & 4.168 & 2.072 \\
\hline 
15 & 5.284 & 2.551 \\
\hline 
20 & 6.224 & 2.969 \\
\hline 
25 & 7.059 & 3.325 \\
\hline 
50 & 10.340 & 4.734 \\
\hline 
75 & 12.866 & 5.808 \\
\hline 
100 & 14.997 & 6.714 \\
\hline 
150 & 18.577 & 8.231 \\
\hline 
200 & 21.596 & 9.510 \\
\hline 
300 & 26.663 & 11.653 \\
\hline 
400 & 30.935 & 13.459 \\
\hline 
500 & 34.700 & 15.050 \\ 
\hline
\end{tabular}
\end{center}
\caption{
\label{tab:means}
\small The table above shows the mean and
standard deviation of $R_n$ for several $n$.}
\end{table}

Orthogonality of the resulting irreps suggests that their
kets might be distinguished by a projector.  Indeed, consider the following
Fredkin gate interferometer.
\begin{center}
\[
\Qcircuit @R=2.0em @C=.55em  {
& \lstick{\ket{0}} & \qw & \gate{H} & \ctrl{1} 
& \gate{H} & \qw & \meter & \cw & \\
&  \lstick{\ket{\psi_{\mathcal{T}_{n-1}}} \otimes \ket{\varphi}} 
& \push{\backslash} \qw & \qw & \gate{U_{(jn)}}  \qwx & \qw & \qw & \qw & \qw
& \rstick{\ket{\psi_{\mathcal{T}_n}}}
}
\]
\end{center}
A reading of $0$ on the classical wire output
implies application of a projector onto the irrep of the
top diagram of Equation \ref{eq:decomposition}, while a reading of 
$1$ on the classical output implies a projector onto the irrep of
the bottom diagram of Equation \ref{eq:decomposition}.  Thus
we obtain the following inductive algorithm for using Fredkin gate
interferometers to project into the irrep of a random Young diagram
on $n$ qubits.

\vbox{
\noindent
\hrulefill

\noindent
{\bf Algorithm for projecting into the irrep of a random Young diagram
of $n$-qubits:}
If $n=1$, do nothing.  Else we suppose $\ket{\psi_{\mathcal{T}_{\ell-1}}}$
within the irrep of an $\ell-1$ qubit Young diagram $\mathcal{T}_{\ell-1}$
and continue as follows:
\begin{enumerate}
\item  
\label{step:one}
Add an $\ell^{\mbox{\footnotesize th}}$ (pure) 
qubit in any state to the system.
\item  
\label{step:two}
If $\mathcal{T}_{\ell-1}$ is rectangular, then form the only possible
$\mathcal{T}_\ell$ and return the system state $\ket{\psi_{\mathcal{T}_\ell}}$.
\item  Else apply a Fredkin gate interferometer to project into an
eigenspace of the SWAP unitary $U_{(j\ell)}$, where
$j$ is the hook-qubit of $\mathcal{T}_{\ell-1}$.
\begin{enumerate}
\item If the eigenvalue is $+1$, then form $\mathcal{T}_\ell$ by
appending a box containing an $n$ on the far right of the first
row of $\mathcal{T}_{\ell-1}$.  
\item  If the eigenvalue is $-1$, then
form $\mathcal{T}_\ell$ by appending the new box below the 
hook-qubit box, i.e. below $j_{q+1}$ per Equation \ref{eq:tableaux}.
\end{enumerate}
\end{enumerate}
The result is a valid $\mathcal{T}_\ell$ and $\ket{\psi_{\mathcal{T}_\ell}}$.
Continue until $\ell=n$.

\noindent
\hrulefill
}

Note that the second paragraph of this section describes why
there is no need to measure in Step \ref{step:two}.  Indeed,
a Young diagram which is rectangular except for one trailing
box at the upper right describes
$[\wedge^{2}(\mathbb{C}^2)]^{\otimes (n-1)/2} \otimes
\mbox{Sym}^1(\mathcal{H}_1)$.  Yet the second factor is
merely $\mathcal{H}_1$.  Thus we are tautologically in the
irrep of this diagram upon adding any extra qubit, regardless
of its state.

\section{Creating e-bits}

\label{sec:ebits}

\begin{figure}[t]
\begin{center}
\includegraphics[scale=0.25]{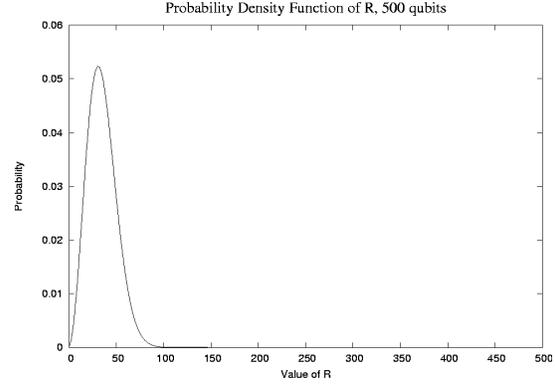}
\end{center}
\caption{
\label{fig:pdf}
\small
Shown above is the probabability density function of $R_{500}$.
This random variable describes how many qubits are not combined
into copies of $\ket{\Psi^-}$ when $(I_{2^n}/2)$ is processed
through the classically-controlled sequence of von Nuemann measurements
of the last section.}
\end{figure}

Recall the Algorithm describing a particular classically-controlled
measurement process in the last section.  Regardless of the initial state
$\ket{\psi}$, some random $\ket{\varphi}$ known to be within the irrep
$\mathcal{L}_{\mathcal{T}}$ of some Young diagram $\mathcal{T}$
always results.  The probability of a particular $\mathcal{T}$
depends on $\ket{\psi}$.  For example, in two qubits a single measurement
is made which projects onto either the singlet
$\ket{\Psi^-}$ or its orthogonal complement, so that the probability
of the resulting diagram $\mathcal{T}$ is 
$|\langle \psi | \Psi^- \rangle|^2$.

It is natural to consider averaging over all $\ket{\psi}$,
equivalently replacing $\ket{\psi}$ by a fully decohered state,
i.e. $\eta=I_{2^n}/2^n=2^{-n} \sum_{j=0}^{2^n-1} \ket{j}\bra{j}$.
In terms of the algorithm, since $\eta=(I_2/2)^{\otimes n}$,
we may equally well consider each added qubit in Step \#1 to
rather be the completely decoherent one-qubit state,
$(1/2)( \ket{0}\bra{0}+ \ket{1}\bra{1})$.  For each Young diagram
$\mathcal{T}$, we also choose an orthonormal basis
$\big\{ \ket{\psi_{j,\mathcal{T}}} 
\big\}_{j=1}^{\mbox{\footnotesize dim} \mathcal{T}}$
and define
\begin{equation}
\rho_{\mathcal{T}} \ = \ (\mbox{dim} \mathcal{T})^{-1}
\sum_{j=1}^{\mbox{\footnotesize dim}\mathcal{T}} 
\ket{\psi_{j,\mathcal{T}}} \bra{\psi_{j,\mathcal{T}}}
\end{equation}
Then $\rho_{\mathcal{T}}$ does not depend on the choice of orthonormal 
basis, and the classically-controlled measurement sequence of the 
algorithm always maps $\eta \mapsto \rho_{\mathcal{T}}$
for some qubit Young diagram $\mathcal{T}$.  In this way, the
sequence of classically-controlled measurements gives rise to 
a random variable on Young diagrams, i.e. by assigning to each
diagram the probability that the process carries the fully
decoherent $\eta$ to a decoherent mixture of the states of the associated
irrep.

Note that $\rho_{\mathcal{T}}$ is pure if and only if
the irrep associated to $\mathcal{T}$ is one-dimensional,
i.e. $\rho_{\mathcal{T}}$ is pure iff the Young diagram
$\mathcal{T}$ is rectangular.  This is well-known in the
two-qubit case.  Namely, if we measure
$\rho_{\mathcal{T}} \in \wedge^2 (\mathcal{H}_1) =
\mbox{span}_{\mathbb{C}} \{ \ket{\Psi^-} \}$, then the measurement
of the completely mixed state has produced a singlet (up to global
phase.)  This may be adjusted by local rotations to produce
an \emph{e-bit} or entangled bit, i.e. $\ket{\Psi^+}$.  In the
$n$-qubit case, each column of the rectangular Young diagram
describes two qubits in which the state must carry a copy
of $\ket{\Psi^-}$, due to Equation \ref{eq:decomposition}.
Hence a rectangular Young diagram shows that $\eta$ has
been converted into a pure state of $n/2$
e-bits.  Furthermore, suppose instead the generic $n$-qubit
Young diagram of Equation \ref{eq:tableaux}.  For convenience,
we label one extra constant $r=n-2q$ for the number of
boxes on the top row of the diagram which do not have a lower
neighbor.  In particular, the irrep is a copy of
$\wedge^2(\mathcal{H}_1)^{\otimes q} \otimes \mbox{Sym}^r(\mathcal{H}_1)$
and so has dimension $r+1$.  For the left hand factor is a copy
of $\mathbb{C}$ while generically the symmetric representation is
spanned by the $\ket{00\ldots 0}$, the singlet, the doublet,
\dots, $\ket{11\ldots 1}$.  With this language set,
the following Observation is an interpretation
of Equation \ref{eq:rec_tens_line}.

\noindent
{\bf Observation:}  Let $\ket{\psi} \in \mathcal{L}_{\mathcal{T}}$ be
arbitrary, for $\mathcal{T}$ a Young diagram per 
Equation \ref{eq:tableaux}.
Then $q=(n-r)/2$ e-bits (in particular copies of
$\ket{\Psi^-}$) may be harvested from $\ket{\psi}$ by pairwise
selecting qubits $j_1$, $k_1$, then $j_2$, $k_2$, through
$j_q, k_q$.  Thus, a similar comment applies to any mixed
$\rho$ supported within $\mathcal{L}_{\mathcal{T}}$,
e.g. the mixture $\rho_{\mathcal{T}}$.

On average, how many e-bits might one harvest from the Algorithm?
Label a random variable $R_n$ to the possible values of $r$
averaged over all Young diagrams according to their probability
of arising from the Algorithm applied to $\eta$.  Then the algorithm 
will produce $Q_n=(n-R_n)/2$ e-bits.  Furthermore, $R_n$ is simpler
to analyze than the random variable on Young diagrams, as follows.
Suppose that after processing $n-1$ qubits, the Algorithm
has created $\rho_{\mathcal{T}}$ for which the value of
$R_{n-1}=d$.  Let $\mathcal{T}_j$, $j=1,2$ be the two possible
new Young diagrams which might result after measuring the eigenvalue
of the swap gate $U_{(jn)}$ for $j$ the hook-qubit.  Then the
irrep decomposition is
\begin{equation}
\mathcal{L}_{\mathcal{T}} \otimes \mathcal{H}_1 \ = \ 
\mathcal{L}_{\mathcal{T}_1} \oplus \mathcal{L}_{\mathcal{T}_2}
\end{equation}
We may form an ensemble for $\rho_{\mathcal{T}} \otimes (I_2/2)$
subordinate to the decomposition at right, i.e. consisting of
a mixture of an orthonormal basis of the irrep of 
$\mathcal{T}_1$ and $\mathcal{T}_2$.  Then the probability
of the two outcomes depends only on the dimension of the two
vector spaces.  Being specific, say $\mathcal{T}_1$ places the
box containing $n$ below the hook qubit, while
$\mathcal{T}_2$ places it to the right.  Then
$\mbox{dim} \mathcal{T}=d+1$,
$\mbox{dim} \mathcal{T}_1=d$, and
$\mbox{dim} \mathcal{T}_2=d+2$.  This provides a rule for
incrementing the subscript of $R_n$ in terms of a Markov
process.
\begin{equation}
\label{eq:probs}
\left\{
\begin{array}{lcl}
\mbox{Prob}(R_n=d+1 | R_{n-1}=d) & = & (d+2)/(2d+2) \\
\mbox{Prob}(R_n=d-1 | R_{n-1}=d) & = & d/(2d+2) \\
\end{array}
\right.
\end{equation}
This recursively defines $R_n$, after we observe that
$\mbox{Prob}(R_1 = 1)=1$.

When evaluating the Markov process numerically, we 
abbreviate $\mbox{Prob}(R_t=j)=P(t,j)$.  By reversing a step,
the recursion rule may be restated as
\begin{equation}
P(t,j) \ = \ \frac{j+1}{2j} P(t-1,j-1) + \frac{j+1}{2j+4} P(t-1,j+1)
\end{equation}
Table \ref{tab:means} then shows some sample means and
deviations of $R_n$, rounded to three decimal places.
Note that the expected value of $R_n$ grows slowly with $n$.
Indeed, for large $d$ Equation \ref{eq:probs} is approximately
a coin toss, corresponding to a random walk in the integers
with expected value equal to its starting point.  Hence, it is
unsurprising that $R_n$ grows slowly, since any increase is
due to the bias between the two probabilities of Equation
\ref{eq:probs}.  This justifies the assertion of the introduction
in that it is cheaper to prepare e-bits by von Neumann measurement
in bulk.  For $R_n$ is the number of $n$-qubits processed
which remain outside e-bits, i.e. within the copy of
$\mbox{Sym}^r(\mathcal{H}_1)$ contained within the irrep of $\mathcal{T}_n$.

A graph of the probability density function of $R_n$ for
$n=500$ qubits appears in Figure \ref{fig:pdf}.  In particular,
suppose that one has a quantum processor capped at 100 qubits,
and one attempts to fuse 500 fully mixed qubits into as many e-bits
as possible using the algorithm.  Any antisymmetric measurement
causes the resulting singlet to be removed.  Then the graph suggests
that one is unlikely to overrun the processor with
this procedure, and only rarely would fewer than $210$ singlet pairs
result.  By Table \ref{tab:means}, we estimate that of $500$ qubits
an average of about $465$ would be fused into e-bits, for a yield of $93\%$.
In contrast, if $n=2$, then the probability that the SWAP measurement
produces an e-bit is $1/4$, for a $25\%$ yeild.

\section{Conclusions and Open Questions}

We have presented a classically controlled algorithm requiring $n-1$
Fredkin gate interferometers, whose placement depends on the outcomes
of earlier classical measurements.  It accomplishes the projection
onto the subspace of $\mathcal{H}_n=(\mathcal{H}_1)^{\otimes n}$ 
described by a random Young diagram, which may
be inferred from the classical measurements.  As an application,
we show that applying this process to a complete mixture
$\rho=I_{2^n}/2^n$ produces a number of entangled bits equal
to the width of the two-row subdiagram.  A large percentage of
the qubits tend to lie in this subdiagram for $n>50$.

A generalization to qutrits or qudits is not obvious.  In this
context, Young diagrams have three (or $d$) rows, and there are
multiple hook qudits.  Distinguishing the location of the new box
is no longer possible with a SWAP, although the irreps are orthogonal
due to their shape.  Thus, the problem would be choosing unitaries
which map the computational basis into the space of each irrep
and demonstrating polynomial size circuits for these unitaries.

In a broader context, this sequence of Fredkin-gate interferometers
appear to be a new application of measurement and classical control,
distinct from stabilizer code checks, quantum teleportation, and
computation with cluster states.  Ongoing work seeks to find other
ways in which measurement gates might be exploited to perform
computations in quantum circuit diagrams, e.g. to observe
spin-flip symmetries.

\noindent
{\bf Acknowledgements:}  The author thanks
RND, AGM, and especially JWA of IDA-CCS for their interest 
in analytically evaluating the Markov process of \S \ref{sec:ebits}.

\appendix

\section{Analytic consideration of the Markov process $R_n$}

Recall $P(t,j)=\mbox{Prob}(R_t=j)$ satisfies the recursion
\begin{equation}
P(t,j) \ = \ \frac{j+1}{2j} P(t-1,j-1) + \frac{j+1}{2j+4} P(t-1,j+1)
\end{equation}
Label $S(t,j)=2^t (j+1)^{-1}P(t,j)$.  Then
\begin{equation}
S(t,j) \ = \ S(t-1,j-1) + S(t-1,j+1)
\end{equation}
This recursion is much simpler to consider analytically.
The base case $P(1,1)=1$ and $P(1,\ast)=0$ else becomes
$S(1,1)=2 (1/2) = 1$ and $S(1,\ast)=0$ else.  The simplified
recursion quickly produces the Table \ref{tab:Stab}, starting the
first row with $t=1$.
The table is most easily analyzed by considering the
(nonzero) diagonals.  By protracted inspection, 
the diagonal terms are differences of binomial coefficients,
where each term of the difference
lies on the same row and they are two apart.  The appropriate
formula is as follows.
\begin{equation}
S(t,j) \ = \ \left\{
\begin{array}{c}
0, \ t-j \mbox{ odd} \\
\left( \begin{array}{c} t-1 \\ (t-j)/2 \\ \end{array} \right) - 
\left( \begin{array}{c} t-1 \\ (t-j-4)/2 \\ \end{array} \right), 
\\
t-j \mbox{ even} \\
\end{array}
\right.
\end{equation}
This formula may be verified by a double induction on $j$ within
$t$.  Hence we have the following for $P(t,j)$.
\begin{equation}
\begin{array}{lcl}
P(t,j) & = & 
2^{-t} (j+1) \times \\
& & \bigg[
\left( \begin{array}{c} t-1 \\ (t-j)/2 \\ \end{array} \right) - 
\left( \begin{array}{c} t-1 \\ (t-j)/2 -2 \\ \end{array} \right)
\bigg] \\
& & \mbox{ if } t-j \mbox{ is even.} \\
\end{array}
\end{equation}

\begin{table}
\begin{center}
\begin{tabular}{rrrrrrrrrrr}
0 & 1 \\
1 & 0 & 1 \\
0 & 2 & 0 & 1 \\
2 & 0 & 3 & 0 & 1 \\
0 & 5 & 0 & 4 & 0 & 1 \\
5 & 0 & 9 & 0 & 5 & 0 & 1 \\
0 & 14 & 0 & 14 & 0 & 6 & 0 & 1 \\
14 & 0 & 28 & 0 & 20 & 0 & 7 & 0 & 1 \\
0 & 42 & 0 & 48 & 0 & 27 & 0 & 8 & 0 & 1 \\
\end{tabular}
\end{center}
\caption{\label{tab:Stab}
\small
This table lists values
of $S(t,j)$, with $t=1$ the top row and $j=0$ the left column.
}
\end{table}

We can now derive an analytic formula for the
expectation of $R_t$.  For convenience, we consider only the case
in which $t$ is even.  Also, we prefer $E(R_t)$ to
$\langle R_t \rangle$.
\begin{equation}
\begin{array}{lcl}
E(R_t) & = & 2^{-t} \sum_{j=0}^{t} j(j+1) S(t,j) \\
& = & 
2^{-t} \sum_{j=0}^{t/2} (2j)(2j+1) \\
& & \bigg[
\left( \begin{array}{c} t-1 \\ t/2 - j  \\ \end{array} \right) - 
\left( \begin{array}{c} t-1 \\ t/2 -j - 2 \\ \end{array} \right)
\bigg] \\
\end{array}
\end{equation}
In order to evaluate the last expression, recall the following summation.
In this equation only, we use $n$ and $j$ out of their context in this
manuscript.
\begin{equation}
\sum_{j=0}^{(n-1)/2} j \left( \begin{array}{c} n \\ j \\ \end{array} \right)
\ = \ n \bigg[ 2^{n-2} - \frac{1}{2}
\left( \begin{array}{c} n-1 \\ (n-1)/2 \\ \end{array} \right) \bigg]
\end{equation}
To verify this, consider the summand.  For the product of $j$
by the appropriate choose is equivalently $n$ times
$n-1$ choose $j-1$, and the rewritten sum now (almost) calculates
the sum of half a row of Pascal's triangle.

Returning to the expression for $E(R_t)$, note the difference of
binomial coefficients
on the RHS produces a telescoping difference up to adjustments.  Collecting
and emphasizing the cancelling $j^2$ terms,
\begin{equation}
\begin{array}{lcl}
2^t E(R_t) &  = &  
6 \left( \begin{array}{c} t-1 \\ t/2-1 \\ \end{array} \right)
+ 20 \left( \begin{array}{c} t-1 \\ t/2 - 2 \\ \end{array} \right) + \\
& & \sum_{j=3}^{t/2}(4j^2-4j^2+16j-12)
\left( \begin{array}{c} t-1 \\ t/2-j \\ \end{array} \right)
\end{array}
\end{equation}
Now one might substitute $j \mapsto t/2-j$ into the index of the
final summation and collect terms appropriately.  Certain
cancellations occur, with the final result being the following:
\begin{equation}
E(R_t) \ = \ \frac{(2t+1)}{2^{t-1}} \left( \begin{array}{c}
t-1 \\ t/2-1 \\ \end{array} \right) -1, \ \mbox{for } t \mbox{ even}
\end{equation}
The result of the similar analysis for $t$ odd is
\begin{equation}
E(R_t) \ = \ \frac{4t}{2^{t}} \left( \begin{array}{c}
t-1 \\ (t-1)/2 \\ \end{array} \right) -1, \ \mbox{for } t \mbox{ odd}
\end{equation}

We briefly remark that the above formula produces asymptotics for
$E(R_t)$ as well.  This is due to a corollary of Stirling's approximation
of the gamma function:
\begin{equation}
\label{eq:stirling}
\sqrt{2 \pi} \; \ell^{\ell+1/2} e^{-\ell} \; < \; \ell ! \; < \; 
\sqrt{2 \pi} \; \ell^{\ell+1/2} e^{-\ell} [1 + (4n)^{-1}]
\end{equation}
Using this to estimate the combination above in terms of factorials
with $t$ even yields
\begin{equation}
E(R_t) \ \approx \ \frac{(2t+1)\sqrt{2}}{ \sqrt{\pi t}}-1, \ t \mbox{ large}.
\end{equation}
Due to Equation \ref{eq:stirling}, the error in the approxation of
the binomial coefficient is roughly $O(t^{-1})$, so that the error
in the above is $O(t^{-1/2})$.  Hence we might simplify further
to $E(R_t) \approx -1+2 \sqrt{2 t / \pi}$.
This has technically only been shown for $t$ even; the odd case
is similar.  The above estimates also argue that
$E(R_t) \in O(\sqrt{t})$ and indeed $E(R_t) \in \Theta(\sqrt{t})$.

\end{document}